\documentclass[doublecol]{epl2}
\usepackage{graphicx}
\usepackage[utf8]{inputenc}
\usepackage{amsmath}
\usepackage{verbatim}
\usepackage{amssymb}
\usepackage{listings}
\usepackage{color}
\newcommand{\expa}[1]{\mathrm{e}^{#1}}   % high exponential groupings a
\newcommand{\expc}[1]{\exp \glc #1 \grc} % low exponential with groupings c
\newcommand{\mean}[1]{\left\langle #1 \right\rangle}
\newcommand{\quot}[1]{``#1''}
\newcommand{\const}{\text{const}}
\newcommand{\VEC}[1]{\mathbf{#1}}
\newcommand{\kvec}{\VEC{k}}
\newcommand{\logb}[2][]{\log^{#1} \glb #2 \grb}  % log-brace,  with - ()
\newcommand{\eq}[1]{eq.~(\ref{#1})}
\newcommand{\fig}[1]{Fig.~\ref{#1}}
\renewcommand\vec[1]{\boldsymbol{#1}}

\newcommand\bigO[1]{\ensuremath{\mathcal{O}({#1})}}
\newcommand{\gld}{\left\{}  % ' group left d' 
\newcommand{\grd}{\right\}}  % ' group right d' 
\newcommand{\glb}{\left(}  % ' group left d' 
\newcommand{\grb}{\right)}  % ' group right d' 
\newcommand{\glc}{\left[}  % ' group left d' 
\newcommand{\grc}{\right]}  % ' group right d' 
\newcommand{\dmax}{d_\text{max}}
\newcommand{\taucorr}{\tau_{\text{corr}}}
\newcommand{\taumix}{\tau_{\text{mix}}}
\begin{document}
\date{\today}\title{Irreversible Markov chains in spin models: 
Topological excitations}

\author{Ze Lei\inst{1}\thanks{\email{ze.lei@ens.fr}} \and Werner Krauth 
\inst{1,2}
\thanks{\email{werner.krauth@ens.fr}}}
\institute{
\inst{1} Laboratoire de Physique Statistique, D\'{e}partement de physique
de l'ENS, Ecole Normale Sup\'{e}rieure, PSL Research University, Universit\'{e}
Paris Diderot, Sorbonne Paris Cit\'{e}, Sorbonne Universit\'{e}s, UPMC
Univ. Paris 06, CNRS, 75005 Paris, France \\
\inst{2} Department of Physics, Graduate School of Science, The University
of Tokyo, 7-3-1 Hongo, Bunkyo, Tokyo, Japan
}

\shortauthor{Z. Lei \etal}


\abstract{
We analyze the convergence of the irreversible event-chain Monte Carlo
algorithm for continuous spin models in the presence of topological
excitations.  In the two-dimensional XY model, we show that the local nature
of the Markov-chain dynamics leads to slow decay of vortex--antivortex
correlations while spin waves decorrelate very quickly.  Using a Fr\'{e}chet
description of the maximum vortex--antivortex distance, we quantify the
contributions of topological excitations to the equilibrium correlations,
and show that they vary from a dynamical critical exponent $z\sim 2$ at
the critical temperature to $z\sim 0$ in the limit of zero temperature.
We confirm the event-chain algorithm's fast relaxation (corresponding to $z=0$)
of spin waves in the harmonic approximation to the XY model.  Mixing times
(describing the approach towards equilibrium from the least favorable initial
state) however remain much larger than equilibrium correlation times at
low temperatures.  We also describe the respective influence of topological
monopole--antimonopole excitations and of spin waves on the event-chain
dynamics in the three-dimensional Heisenberg model.}

\maketitle

\section{Introduction}
\label{s:Introduction}

Classical spin models have played a crucial role in the theory of critical
phenomena and in the formulation of topological phases and their associated
transitions. The analysis of vortices and their interactions in the
two-dimensional XY model has lead, in particular, to the development of the
Kosterlitz--Thouless theory\cite{KosterlitzThouless1973}, which initiated
the era of topology in condensed-matter physics.  Likewise, spin models have
been instrumental in the continued development of the Markov-chain Monte Carlo
method, and especially in the invention of advanced sampling methods.  Cluster
Monte Carlo algorithms \cite{Wolff1989cluster} were of prime importance to show
that Kosterlitz--Thouless theory actually applied to the phase transition in
the two-dimensional XY-model \cite{HasenbuschXY2005}.  Monte Carlo methods also
elucidated the role of topological excitations in other models, such as the
three-dimensional Heisenberg model \cite{HolmJanke1994monte}.

\begin{figure}[htb]
\includegraphics[width=\columnwidth]{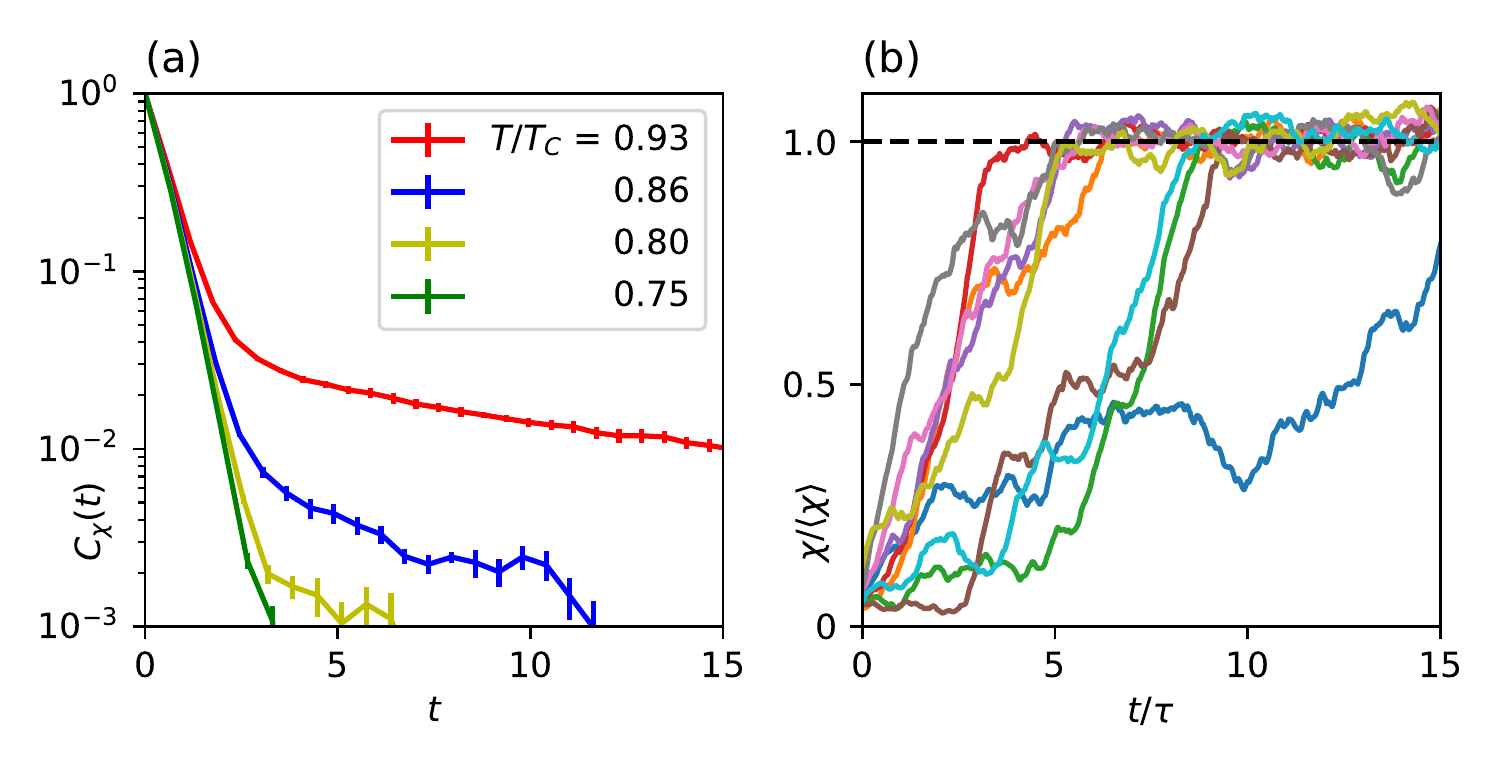}
\caption{Time evolution of the spin susceptibility in the XY model on a
$64\times 64 $ square lattice (time $t$ measured in sweeps). {\it Left: }
Susceptibility autocorrelations below $T_c$. {\it Right: }
Individual susceptibility evolutions at $T/ T_c = 0.93$ starting from random
initial configurations (equilibrium autocorrelation time $\tau$). Large 
sample-to-sample fluctuations are apparent. }
\label{f:XY_Chi}
\end{figure}

In recent years, irreversible Monte Carlo algorithms have increasingly
come into focus. In these methods, the asymptotic steady state (reached
in the long-time limit) still corresponds to thermodynamic equilibrium,
but it is realized with non-zero probability flows.  The event-chain Monte
Carlo algorithm \cite{Bernard2009,Michel2014JCP}, in particular, implements
the global balance condition in a maximally asymmetric way. It relies on the
concept of lifted Markov chains \cite{Diaconis2000}.  Besides short-range and
long-range particle systems \cite{Bernard2011,Kapfer2015PRL,Harland2017}, the
event-chain algorithm applies to continuous spin models such as the 2D and 3D
XY model\cite{MichelMayerKrauth2015,KimuraHiguchi2017} and the 3D Heisenberg
model\cite{Nishikawa2015}. Improved convergence time scales were generally
observed.

In this paper, we discuss the influence of topological excitations and of spin
waves on the convergence of the event-chain algorithm, mostly concentrating on
the two-dimensional XY model with its energy
\begin{equation}
E = - J \sum_{ \langle i, j\rangle} \vec S_i\cdot \vec S_j ,
\label{e:energy_XY}
\end{equation}
with two-dimensional unit spins $\vec S_k = (S_k^x,S_k^y) =
(\cos\phi_k,\sin\phi_k)$ on a square lattice with  $N = L \times L$
sites.  In \eq{e:energy_XY}, the bracket $\langle , \rangle$ denotes
nearest neighbors. For the XY model, the event-chain algorithm (see
\cite{MichelMayerKrauth2015}) rotates a given spin $\vec S_i$ in positive sense
in a sequence of infinitesimal moves until further rotation is vetoed through
the factorized Metropolis algorithm \cite{MichelMayerKrauth2015}. At this
event, spin $i$ comes to a halt, and the neighbor that triggered the veto
takes over, again rotating in positive direction. The event-chain algorithm
violates the detailed-balance condition, but respects global balance.  The
latter is necessary to ensure convergence towards the equilibrium Boltzmann
distribution.  We also consider the harmonic approximation of the XY model
\cite{Wegner1967}, where in the energy of \eq{e:energy_XY} each term $\vec
S_i\cdot \vec S_j = \cos(\phi_i-\phi_j)$ is approximated by $ 1 - \tfrac 12
(\phi_i - \phi_j)^2$,  and, finally, the three-dimensional Heisenberg model,
where the spins $\vec S_i$ are three-dimensional unit vectors.  The XY model
features vortex excitations and it is the unbinding of vortex--antivortex
pairs which takes place at the critical temperature $T_c = 0.893J$. Below the
critical temperature, however, the large-scale excitations of the XY model are
spin waves.  We will argue that the two-stage susceptibility autocorrelation
at low temperature (see \fig{f:XY_Chi}a) corresponds in fact to the fast
decay of spin waves under event-chain dynamics and to the slow decay of the
vortex--antivortex pairs. For $ T/ T_c \to 0$, where vortices are tightly
bound, the event-chain algorithm is asymptotically fast ($z\approx0$),
as we corroborate by simulations. However, the equilibrium correlations
do not give the complete picture of the time behavior of the Markov chain
under consideration. Indeed, one may study the relaxation to equilibrium
after a quench from another temperature (typically from $T = \infty$ to
$T < T_c$). Here, a wide spectrum of relaxation times become relevant, and
equilibration can take much longer than the equilibrium correlation time $\tau$
(see \fig{f:XY_Chi}b). The quench dynamics is sensitive to the mixing time,
which quantifies the approach towards equilibrium from the most unfavorable
initial configuration \cite{Levin2008}. Although the equilibrium correlations
are described by a dynamical critical exponent $z \sim 0$ as $T/ T_c \to 0$,
we will argue that the mixing time remains at $z\sim 2$.

\section{Vortex--antivortex pairs, max-distances}

\begin{figure}[ht]
\includegraphics[width=\columnwidth]{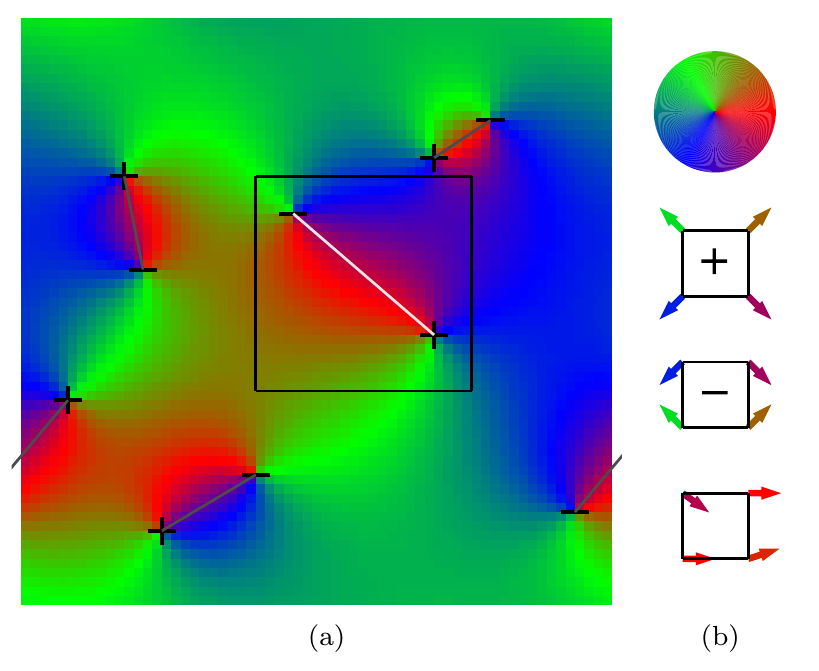}
\caption{Vortices in the XY model. \emph{Left}: Configuration with
$5$ vortices (\quot{+}) and $5$ antivortices (\quot{-}).  The lines
indicate matched vortex--antivortex pairs in the optimal assignment (see
\eq{eq:assignment-min-energy}), and the length of the longest line (shown in
white) equals the max-distance.  A subsystem containing the max-distance pair
is highlighted. \emph{Right, from above}: Color code for the spin orientations,
vortex, antivortex, and neutral plaquette configuration.}
\label{f:vort-match}
\end{figure}

\begin{figure}[ht]
\includegraphics[width=\columnwidth]{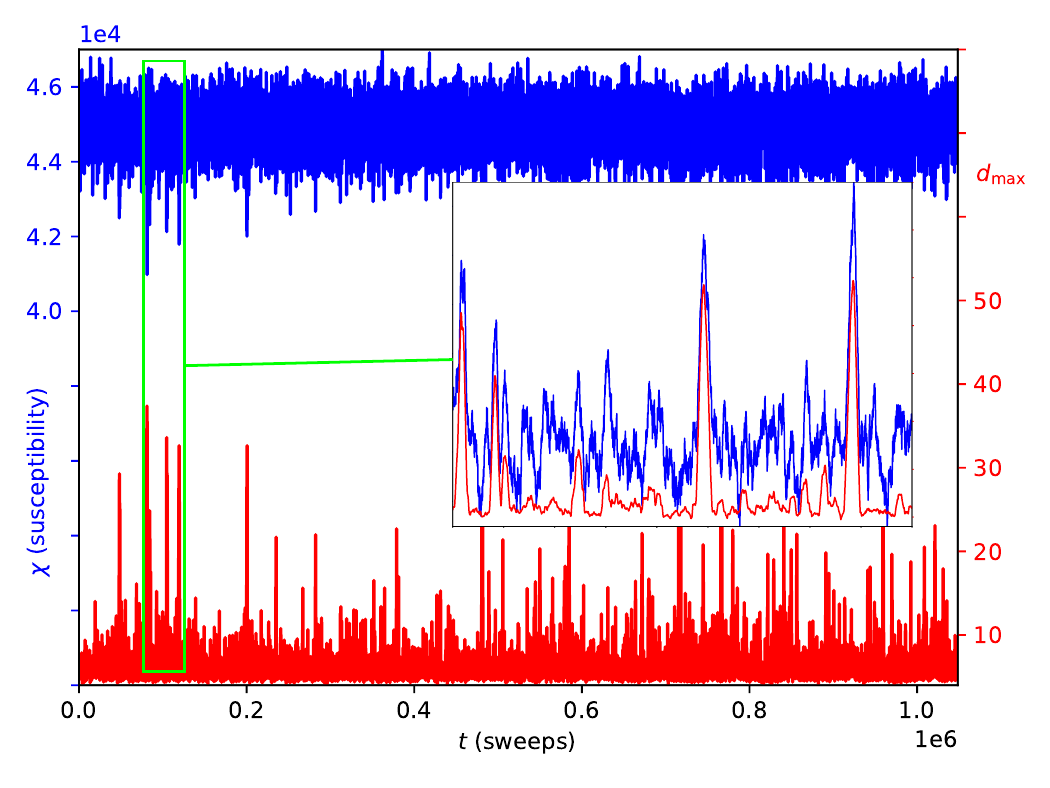}
\caption{Time-evolution of the vortex--antivortex max-distance in
the $384\times384$ XY model at $T/T_c = 0.933$ compared to that of the
susceptibility (smoothed over a small time window). The inset illustrates the
correlation between max-distance and susceptibility in greater detail.}
\label{f:max-d-evolution}
\end{figure}

For the XY model on a square lattice, vortices or antivortices, located
on plaquettes delimited by four spins, are  signalled by differences of
neighboring spins that do not sum to zero when going around the plaquette
in positive sense, but rather to $2\pi$ (vortex) or to $-2\pi$ (antivortex)
(see \fig{f:vort-match}b).  With periodic boundary conditions, vortices
and antivortices appear in pairs.  In a configuration with $n$ such pairs,
the vortices ($v_1, v_2,...,v_n$) can be paired up with the antivortices
($a_{P_1}, a_{P_2},...,a_{P_n}$) according to one of the $n!$ permutations
$P$. We suppose that the physically relevant pairing corresponds to the
minimum of the Kosterlitz--Thouless vortex--antivortex-pair energy $\pi J_R
\logb{R} +  2 E_c$, where the core energy $E_c$ is the same for all
configurations of $n$ pairs and where the value of the renormalized stiffness
$J_R$ of Kosterlitz--Thouless theory does not influence the minimum
\cite{KosterlitzThouless1973}.  We thus neglect interactions between different
vortex--antivortex pairs.  The proper association of each vortex $v_i$ with
its antivortex $a_{P_i}$ defines an assignment problem (see \fig{f:vort-match}a)
aimed at minimizing the objective function $\epsilon$:
\begin{equation}
\label{eq:assignment-min-energy}
 \epsilon(\gld v_i, a_{P_i} \grd = \sum_{i=1}^n \log |\vec{R}_{v_i} - 
\vec{R}_{a_{P_i}}|.
\end{equation} 
The optimal assignment of the $n$ vortex--antivortex pairs can be determined
with standard algorithms \cite{Papadimitriou1982}. Among it, the pair
$(v_i, a_{P_i})$ of largest separation defines the configuration's
\emph{max-distance} $\dmax$.  Remarkably, the time evolution of the
max-distance during a computation mimics that of the susceptibility (see
\fig{f:max-d-evolution}). Large vortex--antivortex pairs (indicated by $\dmax
\gg 1$) and small susceptibilities are particularly well correlated, and both
persist on long time scales (see inset of \fig{f:max-d-evolution}).

We suggest that at low temperature the max-distance length scale determines
the relaxation time scale. To show this, we prepare initial configurations
with only two vortex pairs arranged in a square of length $\dmax = d$ (such
a configuration can be constructed with periodic boundary conditions). We
then track the time needed for the susceptibility to reach the equilibrium
value (within a few percent). At temperature $T \sim T_c$, the system quickly
generates many vortices that screen the distribution of the initial scale. In
contrast, at low temperature, vortex--antivortex pairs at distance $d$ must
approach each other before they can be annihilated.  Indeed, we find that the
time to converge the square-shaped configuration of fixed $d$ is independent of
the system size $L$, and proportional to \bigO{d^2}.  Taking $d = \bigO{L}$,
this implies that the mixing  time $\taumix$ (the time to reach equilibrium
from the most unfavorable initial condition \cite{Levin2008}), is at least
\bigO{L^2}.

For $L \to \infty$, the probability to have a vortex--antivortex pair spaced by
$\vec{d}$ is:
\begin{align}\label{e:LogEnergy}
 P(\vec{d}) &= \frac{1}{Z} \expa{-\beta E_{p}(d)} \nonumber\\
      &= \frac{1}{Z} (d)^{-\pi\beta J_R} \\
      &\propto d^{-\pi\beta J_R}\nonumber ,
\end{align}
where $E_p$ is the pair energy of Kosterlitz--Thouless theory
\cite{KosterlitzThouless1973}.  Because of \eq{e:LogEnergy}, the distribution
of the max-distance for $n$ vortices must be polynomial for $\dmax \to
\infty$. For $T/ T_c \to 0$, the power-law exponent must diverge as
the vortex--antivortex pairs are more and more tightly bound.

\section{Fr\'{e}chet distribution, vortex max-distance}

\begin{figure}[ht]
\begin{center}
\includegraphics[width=1.0\columnwidth]{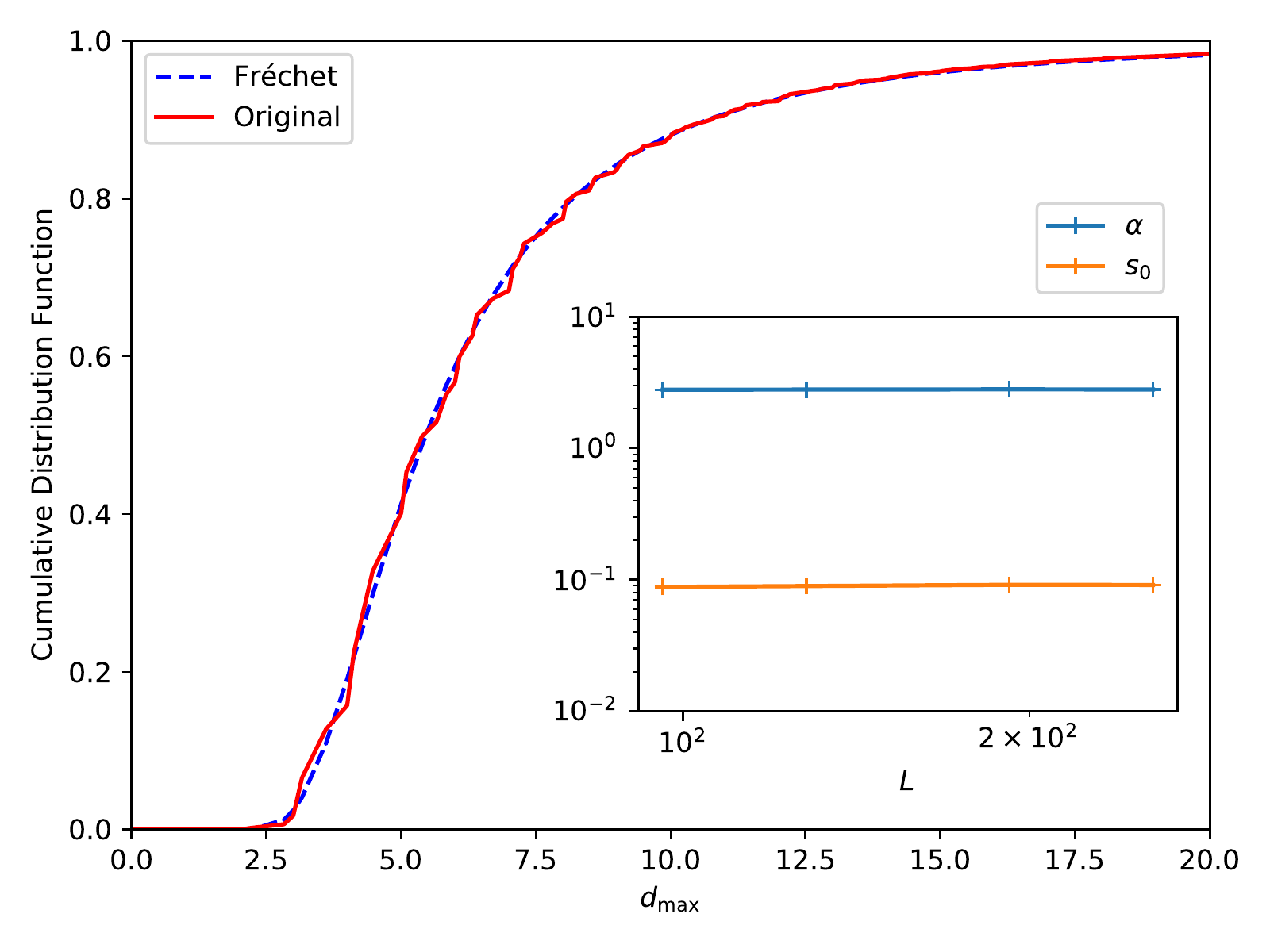}
\caption{Cumulative max-distance distribution in the XY model at $L= 256$ and
$T/T_c = 0.965$ fitted with the Fr\'{e}chet distribution with $\alpha=2.80$ and
$s=4.79$ ($s_0 =  0.0913$).  The inset illustrates that the fitting parameters
$\alpha$ and $s_0$ are independent of $L$.}
\label{f:Frechet_scaling_beta_alpha}
\end{center}
\end{figure}

At temperatures below $T_c$, for $L \to \infty$, vortex--antivortex pairs
are bound \cite{KosterlitzThouless1973}, so that the equilibrium max-distance
$\dmax$ is much smaller than the system size, and its probability distribution
$p(\dmax)$ decays algebraically for large arguments (see \eq{e:LogEnergy}).
The $L\times L$ system can be divided into $n^2$ practically independent
subsystems of size $L/n \times L/n$. The max-distance of the large system at
scale $L$ is the maximum of $n^2$ independent max-distances on a scale $L/n$.
Extreme-value statistics \cite{de2007extreme} allows one to connect the
distribution $p(\dmax)$ at scale $L$ with the one at $L/n$. It must correspond
to the Fr\'{e}chet distribution (with zero minimum value), specifically:
\begin{align}\label{eq:frechet_distribution}
 p(\dmax) =   
  \frac{\alpha}{s}\glb \frac{\dmax}{s} \grb^{-1 - \alpha} 
 \expc{-\glb \frac{\dmax}{s}  \grb^{-\alpha}}\\
 \intertext{with its cumulative distribution}
  P(\dmax) = \expc{ - \glb \frac{\dmax}{s} \grb ^{-\alpha}}.
\end{align}
Here, $\alpha$ describes  the power-law decay of the max-distance distribution
for large arguments (which is the same on scales $L$ and $L/n$), and $s$  
sets its
$L$-dependent scale.  The maximum of $n$ independent samples of a Fr\'{e}chet
distribution with parameters $(\alpha, s)$ is distributed following a
Fr\'{e}chet distribution with parameters $(\alpha, n^{1/\alpha} s)$. It
then follows that the Fr\'{e}chet distribution of the max-distance in a system
of size $L$ must be described by parameters $(\alpha, L^{2 / \alpha}
s_0)$, where both $\alpha$ and $s_0$ depend on $\beta$, but not on $L$, for
large $L$.  Slightly below $T_c$ already, the Fr\'{e}chet distribution provides
an excellent fit for the max-distance distribution and the fitting parameters
$\alpha$ and $s_0$ are indeed independent of $L$ for a given temperature (see
\fig{f:Frechet_scaling_beta_alpha}).  Also, we note that for $\alpha =2$,
the distribution of $\dmax$ scales with $s \propto L$. This is observed for
$T / T_c \to 1^-$. At low temperatures, we observe $\alpha \propto 1/T$, in
agreement with \eq{e:LogEnergy}.

Below $T_c$, the probability distribution of $\dmax$ scales as $\sim
L^{2/\alpha} \ll L$, as $\alpha > 2$, and we expect the equilibrium correlation
time to scale with $s^2 = L^{4/ \alpha} s_0^2 $:
\begin{equation}
\taucorr^{\text{vortex}} \sim L^{4 / \alpha} \sim 
\begin{cases}
L^2 \quad & \text{for $ T \to T_c ^{-}$}\\
L^{\text{const} T} & \text{for $T \to 0$}
\end{cases} .
\label{e:taucorr_scales}
\end{equation}
The effective dynamical scaling parameter $z = 4 / \alpha$ of the event-chain
algorithm is thus connected to the scale parameter of a Fr\'{e}chet
distribution and predicted to vanish in the zero-temperature limit.

\section{Harmonic model, spin waves}

The ansatz of \eq{e:taucorr_scales} for the equilibrium correlations only
describes the relaxation of topological excitations, parametrized by the
max-distance. We now consider spin waves which, below $T_c$, are the dominant
large-distance excitations for local Monte Carlo dynamics, where they take
\bigO{L^2} sweeps to relax.  In the event-chain algorithm, they relax in
\bigO{L^0} sweeps, so that our ansatz is indeed consistent.  To show this, we
study the harmonic model, an approximation to the XY Hamiltonian, where the
spin variables $\phi$ interact as follows:
\begin{equation}
E = \frac J 2\sum_{  \langle i, j\rangle } (\phi_i - \phi_j)^2. 
\end{equation}
This model is exactly solved by taking Fourier modes as the independent
variables \cite{Wegner1967}.  The two-dimensional harmonic model has
algebraically decaying spin correlations with an exponent that approaches zero
as $T/ T_c \to 0$.  From the exact solution of the harmonic model, it follows
that the difference of $\phi$ on sites distant by \bigO{L} is on a scale
\begin{align}
\sigma^{\text{eq}}(L) \propto
  \begin{cases}
    \sqrt{L}  & \quad \text{ if } d = 1 \nonumber\\
    \sqrt{\log L}  & \quad \text{ if } d = 2\\
    1 & \quad \text{ if } d \geq 3\nonumber
  \end{cases}.
\end{align}

\begin{figure}[ht]
\includegraphics[width=\columnwidth]{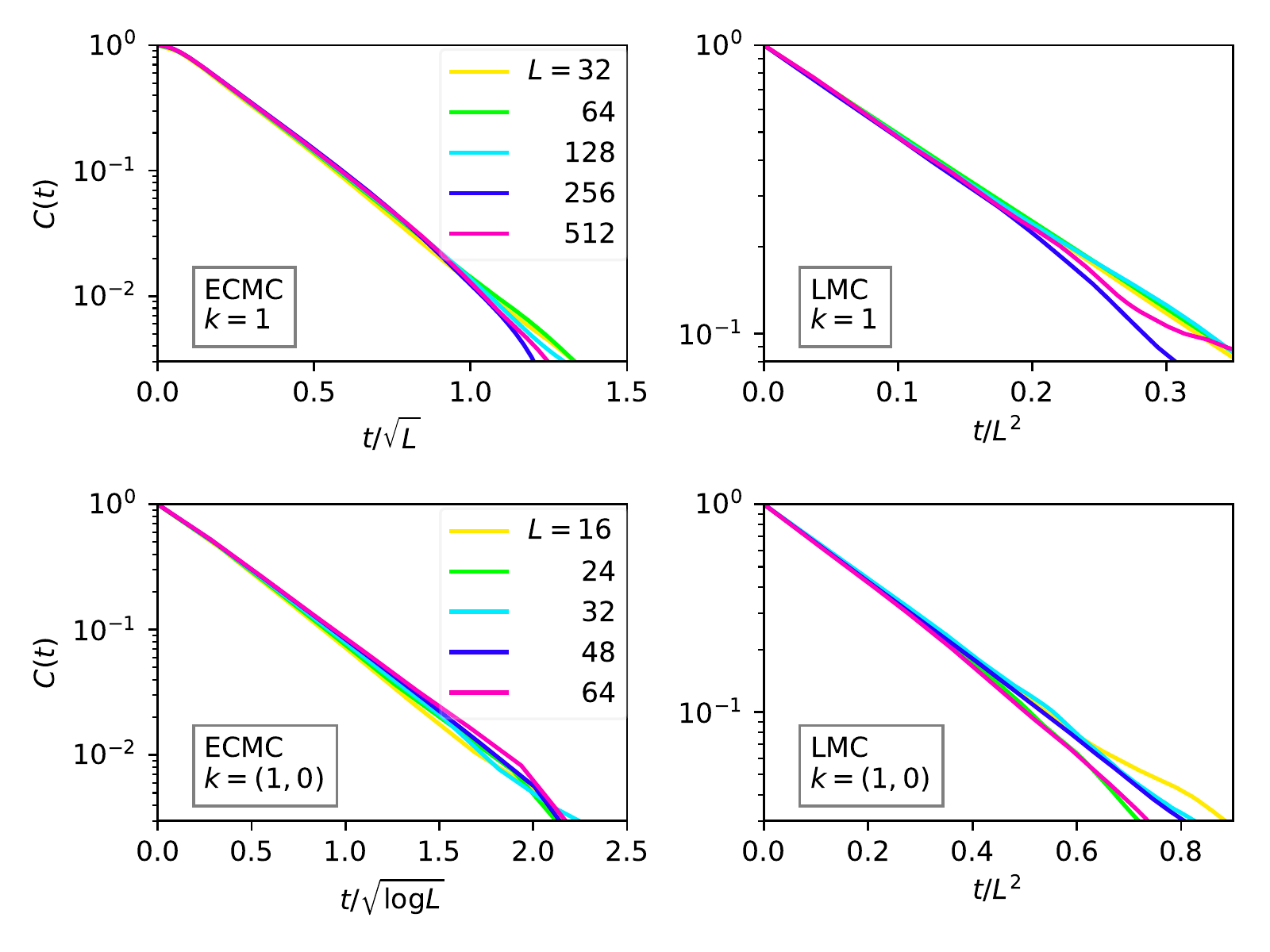}
\caption{Equilibrium auto-correlation functions $C(t)$ of the lowest Fourier 
modes in the harmonic model for the event-chain algorithm (ECMC) and for local 
Monte Carlo (LMC).
\emph{Upper}: $C(t)$ for the Fourier mode $k=1$ in 1D.
\emph{Lower}: $C(t)$ for the Fourier mode $k=(1,0)$ in 2D.
Data are in agreement with the scaling of \eq{taucorr_harmonic}.
}
\label{f:ECMC_harmonic}
\end{figure}

The event-chain algorithm for the harmonic model can only increase the value
of $\phi_i$.  We find that in one sweep (\bigO{N} events), the mean value
$\mean{\phi_i}$ of a configuration increases by \bigO{1}.  The correlation time
of the algorithm is reached when the mean increase per site is on the order of
the equilibrium correlation $\sigma^{\text{eq}}$.  This implies the relation
\begin{equation}
 \taucorr^{\text{harm}}  \sim \sigma^{\text{eq}}(L).
 \label{taucorr_harmonic}
\end{equation}
The \eq{taucorr_harmonic} predicts a dynamical scaling exponent of $1/2$ for
the $1D$ harmonic model, and an exponent $z=0$ in higher dimensions. This fast
dynamical scaling, in sharp contrast to the behavior of the local Metropolis
algorithm (with $z \sim 2$) is verified for the autocorrelation times for
Fourier modes with small $\kvec$ (see \fig{f:ECMC_harmonic}).

In the XY model below $T_c$, the two types of excitations generate two
time scales for the equilibrium autocorrelation function of the event-chain
algorithm. This corresponds to what is observed in the susceptibility, where we
thus associate the fast initial decay with spin waves ($\taucorr^{\text{harm}}
\sim \const$), and the slow decay with topological excitations
(vortex--antivortex pairs, $\taucorr^{\text{vortex}} \sim L^ {\const T}$ at
low temperature and $\taucorr^{\text{vortex}} \sim L^2$ for $T/ T_c \to 1^-$)
(see \fig{f:XY_Chi}).

\section{Monopoles, Bloch waves}

Topological excitations also play a prominent role in other spin models, for
example the 3D Heisenberg model. Low-temperature excitations in that model can
also be described by spin waves in addition to topological excitations. Spin
waves again come with a dynamical critical exponent $\sim 0$.  Heisenberg-model
monopoles and anti-monopoles are again located on the dual lattice, and they
can be identified using a well-defined algorithm \cite{Berg1981}.

In the 3D Heisenberg model, monopole--antimonopole pairs proliferate near the
critical point\cite{HolmJanke1994monte}. Their excitation energy increases with
the separation $d$ as  \bigO{d} \cite{Lau1988,Ostlund1981}.  The event-chain
algorithm, at low temperature, again moves each monopole or antimonopole by
$\bigO{1}$ per sweep. From initial configurations with pairs separated on a
scale $\bigO{L}$, we find that relaxation towards equilibrium takes $\bigO{L}$
sweeps (rather than \bigO{L^2}, as for the XY model).  Configurations with
widely separated pairs play no role at low temperature, and the spin waves are
again treated efficiently in the event-chain algorithm, so that $z \to 0$
for $T/ T_c \to 0$.  Nevertheless, the mixing time scale for the approach to
equilibrium from an unfavorable configuration is $\bigO{L}$ sweeps.

Finally, there are other types of topological excitations, besides the
point-like ones (vortices, monopoles) discussed here. Bloch modes, in the XY
model with periodic boundary conditions, correspond for example to a state in
which the spins rotate by $2\pi$ as one coordinate, say $x$, goes from $0$ to
$L$. Bloch waves are a slow mode in the event-chain algorithm for the XY model
(but not in the Heisenberg model), and  stable on a time scale \bigO{L^2} at
low temperature in both 2D and 3D.

\section{Conclusions}

In this paper, we exhibited a considerable speed-up for the relaxation
of spin-wave excitations of the event-chain algorithm compared to the
local Monte Carlo algorithm. Indeed, in the harmonic model, which has only
spin waves, the event-chain algorithm equilibrates in a constant number of
sweeps for $d >1$, whereas the local algorithm equilibrates with $z \sim 2$.
We have also studied the relaxation of topological excitations, namely the
vortex--antivortex pairs in the 2D XY model and the monopole--antimonopole
pairs in the Heisenberg model. In the XY model, below the critical temperature,
vortex--antivortex pairs are bound, and we parametrize this binding with
a single parameter, the max-distance $\dmax$ that can be computed with a
combinatorial-optimization algorithm. We find that the probability distribution
of $\dmax$ is a Fr\'{e}chet distribution (with zero minimum value).  In the XY
model, the vortex--antivortex potential is very weak, leading to a $\dmax^2$
relaxation time and, at worst, an $L^2 $ mixing time. However, 
equilibrium-correlation time scales are much smaller. In the event-chain 
algorithm,
these vortex--antivortex excitations are no longer concealed by the spin
waves, and they in fact constitute the slowest dynamical modes for the
event-chain algorithm.  It is thus found to have a smaller dynamical exponent
than the local Monte Carlo algorithm for all temperatures below $T_c$. In
particle systems, we likewise expect the fast relaxation of phonon modes
(which, in analogy to the spin waves of this paper, are also described by
a harmonic model) to be key to the success of the event-chain algorithm
at high densities \cite{Bernard2011}. However, the fundamental difference
between mixing times (needed to reach equilibrium from the most unfavorable
initial condition) and equilibrium correlation times (needed to move to a new
independent configuration from an equilibrium starting configuration) appears
clearly \cite{Levin2008}. It will certainly have to be taken into account in
applications.

\begin{acknowledgments}
We thank Youjin Deng for helpful discussions in the initial 
stages of this work and Cris Moore for useful suggestions.
\end{acknowledgments}

\bibliographystyle{eplbib.bst}%abbrv}
\bibliography{General.bib}\addcontentsline{toc}{section}{Literaturverzeichnis
}
\newpage

\renewcommand\refname{ }
\end{document}